\documentclass[a4paper,12pt]{article}
\input cohmacros.sty

\advance\textheight3.9cm        
\advance\topmargin-2.3cm
\advance\textwidth2.6cm
\advance\evensidemargin-1.6cm
\advance\oddsidemargin-1.6cm

\begin{document}

\begin{center}
{\LARGE \bf Born's rule and measurement} \\

\vspace{1cm}

\centerline{\sl {\large \bf Arnold Neumaier}}

\vspace{0.5cm}

\centerline{\sl Fakult\"at f\"ur Mathematik, Universit\"at Wien}
\centerline{\sl Oskar-Morgenstern-Platz 1, A-1090 Wien, Austria}
\centerline{\sl email: Arnold.Neumaier@univie.ac.at}
\centerline{\sl \url{http://www.mat.univie.ac.at/~neum}}

\end{center}


\hfill December 20, 2019

\vspace{0.5cm}

\bigskip
\bfi{Abstract.}
Born's rule in its conventional textbook form applies to the small
class of projective measurements only. It is well-known that a
generalization of Born's rule to realistic experiments must be phrased
in terms of positive operator valued measures (POVMs). This
generalization accounts for things like losses, imperfect measurements,
limited detection accuracy, dark detector counts, and the simultaneous
measurement of position and momentum.

Starting from first principles, this paper gives a self-contained,
deductive introduction to quantum measurement and Born's rule, in its 
generalized form that applies to the results of measurements described 
by POVMs. It is based on a suggestive definition of what constitutes a 
detector, assuming an intuitive informal notion of response. 

The formal exposition is embedded into the context of a variaety of 
quotes from the literature illuminating historical aspects of the 
subject. The material presented suggests a new approach to introductory 
courses on quantum mechanics.

\vfill
For the discussion of questions related to this paper, please use
the discussion forum \\
\url{https://www.physicsoverflow.org}.

\vfill
MSC Classification (2010): primary: 81P15, secondary: 81-01

\newpage
\tableofcontents 

\newpage
\section{The measurement process}\label{s.measP}

\nopagebreak
\hfill\parbox[t]{10.8cm}{\footnotesize

{\em We have developed a pure geometry, which is intended to be
descriptive of the relation-structure of the world. The
relation-structure presents itself in our experience as a physical world
consisting of space, time and things. The transition from the
geometrical description to the physical description can only be made by
identifying the tensors which measure physical quantities with tensors
occurring in the pure geometry; and we must proceed by inquiring first
what experimental properties the physical tensor possesses, and then
seeking a geometrical tensor which possesses these properties by virtue
of mathematical identities. \\
If we can do this completely, we shall have constructed out of the
primitive relation-structure a world of entities which behave in the
same way and obey the same laws as the quantities recognised in physical
experiments. Physical theory can scarcely go further than this.}

\hfill Arthur Eddington 1930 \cite[p.222]{Edd}
}

\bigskip

\nopagebreak
\hfill\parbox[t]{10.8cm}{\footnotesize

{\em The following 'laboratory report' of the historic Stern-Gerlach
experiment stands quite in contrast to the usual textbook 'caricatures'.
A beam of silver atoms, produced in a furnace, is directed through an
inhomogeneous magnetic field, eventually impinging on a glass plate.
[...]
Only visual measurements through a microscope were made. No statistics
on the distributions were made, nor did one obtain 'two spots' as is
stated in some texts. The beam was clearly split into distinguishable
but not disjoint beams. [...]
Strictly speaking, only an unsharp spin observable, hence a POV measure,
is obtained.}

\hfill Busch, Grabowski and Lahti, 1995 \cite[Example 1, p.7]{BusGL}
}

\bigskip

\nopagebreak
\hfill\parbox[t]{10.8cm}{\footnotesize

{\em
If you visit a real laboratory, you will never find there Hermitian
operators. All you can see are emitters (lasers, ion guns, synchrotrons
and the like) and detectors. The experimenter controls the emission
process and observes detection events. [...]
Traditional concepts such as ''measuring Hermitian operators'', that
were borrowed or adapted from classical physics, are not appropriate
in the quantum world. In the latter, as explained above, we have
emitters and detectors, and calculations are performed by means of
POVMs.}

\hfill Asher Peres, 2003 \cite[p.1545f]{Per2003}
}

\bigskip

All traditional foundations of quantum mechanics depend heavily  -- far
too heavily -- on the concept of (hypothetical, idealized) measurements
{\em exactly} satisfying Born's rule, a nontrivial technical rule far
from being intuitive. This -- almost generally assumed -- exact validity
without a precise definition of the meaning of the term measurement
is probably the main reason why, nearly 100 years after the discovery
of the basic formal setting for modern quantum mechanics, these
foundations are still unsettled. No other scientific theory has such
controversial foundations.

The source of this poor state of affairs is that Born's rule for
projective measurements, the starting point of the usual
interpretations, constitutes a severe idealization of measurement 
processes in general. Except in a few very simple cases, it is too far 
removed from experimental practice to tell much about real measurement, 
and hence about how quantum physics is used in real applications. But 
foundations that starts with idealized concepts only do not provide a 
safe ground for interpreting reality. 

More general theoretical descriptions of quantum measurements were
introduced in 1970 by \sca{Davies \& Lewis} \cite{DavL}, 45 years after
Heisenberg's 1925 paper initiating modern quantum physics.  A very
readable account was given 4 years later by \sca{Ali \& Emch}
\cite{AliE.meas}. Since this paper appeared, another 45 years passed.
These general measurement schemes are based on a \bfi{discrete POVM}
(also called a \bfi{discrete resolution of the identity}),
a family of finitely many Hermitian positive semidefinite operators
$P_k$ on a Hilbert space summing to 1,
\[
\sum_k P_k=1.
\]
This generalization accounts for things like losses, imperfect
measurements, limited detection accuracy, dark counts, and the
simultaneous measurement of position and momentum. POVMs were soon found
useful for concrete applications to the calibration of quantum systems
(\sca{Helstrom} \cite{Hel}). \sca{Brandt} \cite{Bra} gives a short,
more substantive history. For a fairly concise, POVM-based
exposition of the foundations of quantum mechanics see, e.g.,
\sca{Englert} \cite{Eng}.

POVMs are indispensable in quantum information theory. Indeed, the
well-known textbook by \sca{Nielsen \& Chuang} \cite{NieC} introduces
them even before defining the traditional projective measurements.
POVMs are also needed to describe quite ordinary experiments without
making the traditional textbook idealizations. For example, the original
Stern-Gerlach experiment did (in contrast to its textbook caricature)
not produce two well-separated spots on the screen but two overlapping
lips of silver. This outcome cannot be described in terms of a
projective measurement but needs POVMs.

Similarly, joint measurements of position and momentum, which are
ubiquitous in engineering practice, cannot be described in terms of a
projective measurement. Born's rule in the pre-1970 form does not even
have idealized terms for these.

Books featuring POVM measurements include
\sca{Busch} et al. \cite{BusLM,BusGL,BusL2,BusLPY},
 \sca{de Muynck} \cite{deMuy2002},
\sca{Holevo} \cite{Hol1982,Hol2001,Hol2012},
\sca{Nielsen \& Chuang} \cite{NieC} and \sca{Peres} \cite{Peres}.

In foundational studies, the projective idealization, taught in
nearly \footnote{
see, e.g., \sca{Neumaier} \cite{Neu.7basic} and Footnote
${}^{\ref{f.proj}}$
} 
every textbook on quantum mechanics, is traditionally taken far too
seriously. It counts as the indisputable truth about everything measured
on the most fundamental level, to which everyone pays lipservice.
But from the practical point of view, this idealization seems to be only
a didactical trick to make the formal definitions of quantum mechanics
easier to swallow for the newcomer -- without needing seemingly abstract
notions such as POVMs. On the other hand, Born's rule for projective
measurement needs spectral theory up to the spectral theorem, itself not
a simple subject. Thus there seems room for improvement even on the
didactical level.

The purpose of this paper is to give an intuitive, self-contained
approach to quantum measurement in the spirit of Eddington. 
Born's rule is not postulated as the starting point, but appears as a
derived statement valid under the conditions specified in its
derivation. This sheds a new light on the understanding of Born's rule
and eliminates the most problematic features\footnote{
All problematic features known to me are collected in my recent book
(\sca{Neumaier} \cite[Section 14.3]{Neu.CQP}), following the preprint
\sca{Neumaier} \cite[Section 3.3]{Neu.Ifound}.
} 
of its uncritical use.

We do not discuss the question of whether (or which form of) state
reduction occurs when subjecting a system to measurement. Neither do
we discuss the measurement problem, i.e., how Born's rule can be
justified in terms of a unitary dynamics of a larger quantum model
containing a measured system and a measuring detector. For such quantum
models of POVM measurements see \sca{Busch \& Lahti} \cite{BusL2} and
several chapters in \sca{Busch} et al. \cite{BusGL, BusLPY}.

In the remainder of this section we give motivations and precise 
definitions of states (positlive linear functionals) and quantum 
detectors (consisting of a POVM and a scale), and give a precise 
specification of what is measured by such a detector, covering the most 
general case. Section 2 gives numerous examples before considering
the special case singled out by tradition -- that of projective 
measurements. The section ends with proposing a new design for 
introductory courses on quantum mechanics, based on the preceding.
Section \ref{s.unc} gives a thorough, precise discussion of various 
aspects of uncertainty in quantum measurements and relates them to the 
thermal interpretation of quantum physics as defined in my recent book
{\em Coherent quantum physics} (\sca{Neumaier} \cite{Neu.CQP}).

\subsection{States and their properties}\label{ss.states}

\nopagebreak
\hfill\parbox[t]{10.8cm}{\footnotesize

{\em
Quantum mechanics tells us that whatever comes from the emitter is
represented by a state $\rho$ (a positive operator, usually normalized
to 1).}

\hfill Asher Peres, 2003 \cite[p.1545]{Per2003}
}

\bigskip

We motivate the formal setting of this paper by considering the
polarization of classical light, as in Section 8.6 of \sca{Neumaier}
\cite{Neu.CQP}, from which a few paragraphs are taken.

A ray (quasimonochromatic beam) of classical polarized light of fixed
frequency is characterized by a state, described (\sca{Mandel \& Wolf}
\cite[Section 6.2]{ManW}) by a real \bfi{Stokes vector}
\[
S=(S_0,S_1,S_2,S_3)^T={S_0\choose \Sb}
\]
with
\[
S_0\ge |\Sb| = \sqrt{S_1^2+S_2^2+S_3^2},
\]
The Stokes vector is a classical observable vector quantity.
Equivalently, the state can be described by a \bfi{coherence matrix}, a
complex positive semidefinite
$2\times 2$ matrix $\rho$. These are related by
\[
\rho = \half(S_0 + \Sb \cdot\Bsigma)
=\half\pmatrix{S_0+S_3 & S_1-iS_2\cr S_1+iS_2 & S_0-S_3},
\]
where $\Bsigma$ is the vector of Pauli matrices
$\sigma_1,\sigma_2,\sigma_3$. If we define $\sigma_0:=1$ as the identity
operator we have
\[
S_k=\<\sigma_k\> \for k=0,\ldots,3,
\]
where
\[
\<X\>:=\Tr\rho X
\]
denotes the \bfi{q-expectation} of the matrix $X\in\Cz^{2\times 2}$.
In particular, $\Tr\rho=\<1\>=S_0$ is the \bfi{intensity} of the beam.

The quotient $p=|\Sb|/S_0\in[0,1]$ is the \bfi{degree of polarization}.
Since
\[
\det\rho=(S_0^2-S_3^2)-(S_1^2+S_2)^2=S_0^2-\Sb^2,
\]
the fully polarized case $p=1$, i.e., $S_0=|\Sb|$, is equivalent to
$\det\rho=0$, hence holds iff the rank of $\rho$ is $0$ or $1$. In this
case we say that the state is \bfi{pure}. Thus the pure states
correspond precisely to fully polarized beams. In a pure state, the
coherence matrix can be written in the form $\rho=\psi\psi^*$
with a \bfi{state vector} $\psi\in\Cz^2$ determined up to a phase, and
the intensity of the beam is
\[
S_0=\<1\>=|\psi|^2=\psi^*\psi.
\]
Thus notions related to a complex Hilbert space of dimension 2 model
the simplest quantum phenomenon: A positive definite Hermitian $\rho$
describes the state of an arbitrary source, the trace of $\rho$ is the
intensity of the source, and certain Hermitian operators represent key
quantities. Note the slight difference to density operators, where the
trace is required to be one.

Combining two independent sources leads to the addition of the
intensities and hence the corresponding densities. Similarly, changing
the intensity amounts to a scalar multiplication of the corresponding
densities. This provides experimental support
for the linearity of typical detector responses, a feature observed in
broad generality.

We may therefore generalize from the polarization experiments to
experiments involving arbitrary quantum systems. Probably having
mastered the above lets every beginning student of quantum physics
accept the generalization. It is enough to say that in nearly
100 years of experimental work it was established beyond reasonable
doubt that not only photon polarization but an arbitrary quantum
system is describable in terms of an arbitrary complex Hilbert space,
with a positive definite Hermitian $\rho$ with finite trace describing
the state of an arbitrary source, the trace of $\rho$ defining the
macroscopic intensity of the source, and certain Hermitian operators
(with details depending on the quantum system) define key quantities.
While this is not a proof, one may refer to authorities when creating
foundations for a beginners course, and only make plausibility arguments
that are easily grasped.

In the following we develop the time-independent part of the formal core
of quantum mechanics -- conventional shut-up-and-calculate quantum
mechanics in a complex Hilbert space $\Hz$, but without Born's rule.
Thus we avoid the traditional a priori link to experimental practice via
an ill-defined notion of measurement. Instead, such a link will be
established through careful definitions.

We describe a \bfi{source} (one of the emitters in the quote by Peres)
by a positive semidefinite Hermitian
\bfi{density operator} $\rho\in \Lin \Hz$; the source is called
\bfi{pure} if the density operator has rank 1, and hence is given by
$\rho=\psi\psi^*$ for some \bfi{state vector} $\psi$. The \bfi{state} of
the source is the positive linear mapping $\<\cdot\>$ that assigns to
each $X\in \Lin \Hz$ its \bfi{q-expectation}\footnote{
Following the convention of \sca{Allahverdyan} et al. \cite{AllBN2},
we add the prefix ''q-'' to all theoretical quantum notions that
suggest by their name a statistical interpretation and hence might
confuse the borderline between theory and measurement.
} 
\lbeq{e.qEx}
\ol X=\<X\>:=\Tr\rho X.
\eeq
More generally, given a fixed state, the \bfi{q-expectation}\footnote{
Traditionally, the q-expectation $\<X\>$ of an operator $X$ is called
the expectation value of $X$. But when $X$ is not normal (and in
particular when it is defective, hence has not even a spectral
resolution), a statistical interpretation in the traditional sense is
impossible. Note that nonnormal q-expectations appear routinely in
quantum field theory, e.g., in the definitions of so-called $N$-point
functions.
It is interesting to note that Dirac's 1930 book, which introduced the
name ''observable'' for operators, used this terminology for arbitrary
linear operators (\sca{Dirac} \cite[p.28]{Dir1}). Later editions make
the restriction to Hermitian (and implictly self-adjoint) operators.
} 
of a vector $X\in(\Lin \Hz)^m$ with operator components is the vector
$\ol X=\<X\>\in\Cz^m$ with components $\ol X_j=\<X_j\>$. Its
\bfi{q-uncertainty} is the nonnegative number
\lbeq{e.sigmaXv}
\sigma_X:=\sqrt{\<(X-\ol X)^*(X-\ol X)\>}=\sqrt{\<X^*X\>-|\ol X|^2}.
\eeq
We may also define the \bfi{q-covariance matrix}
\[
C_X:=\<(X-\ol X)(X-\ol X)^*\>\in\Cz^{n\times n},
\]
in terms of which
\[
\sigma_X=\sqrt{\tr C_X}.
\]
The uncertainty relation (due to \sca{Robertson} \cite{Rob29}), which
asserts that for non-commuting Hermitian operators $A,B\in\Lin\Hz$,
\lbeq{e.uncRel}
\sigma_A\sigma_B\ge \half |\<[A,B]\>|,
\eeq
follows from the definitions.\footnote{
Indeed, the relation remains unchanged when subtracting from $A$ and
$B$ its q-expectation, hence it suffices to prove it for the case
where both q-expectations vanish. In this case, $\<A^2\>=\sigma_A^2$ and
$\<B^2\>=\sigma_B^2$, and the Cauchy--Schwarz inequality gives
$|\<AB\>|^2\le \<A^2\>\<B^2\>=\sigma_A^2\sigma_B^2$, hence
$|\<AB\>|\le\sigma_A\sigma_B$. On the other hand, one easily checks that
$i\im\<AB\>=\half\<[A,B]\>$, so that
$\half |\<[A,B]\>|=|\im\<AB\>|\le |\<AB\>|$. Combining both inequalities
gives the assertion.
} 
The q-expectations and q-uncertainties are complex numbers or vectors
providing theoretical properties of the state, independent of any notion
of measurement. The uncertainty relation also does not involve the
notion of measurement. Thus these notions belong to the formal core of
quantum mechanics. Thus, a priori, no statistical implication is
assumed.

The collection of all q-expectations completely determines the state.
Hence any property of the state can be expressed as a function of
q-expectations.

\subsection{Detectors, scales, and POVMs}\label{ss.detectors}

\nopagebreak
\hfill\parbox[t]{10.8cm}{\footnotesize

{\em
The theorist's problem is to predict the probability of response of this
or that detector, for a given emission procedure. Detectors are
represented by positive operators $E_\mu$, where $\mu$ is an arbitrary
label whose sole role is to identify the detector. The probability that
detector $\mu$ be excited is $\Tr(\rho E_\mu)$. A complete set of
$E_\mu$, including the possibility of no detection, sums up to the unit
matrix and is called a positive operator valued measure (POVM).
\\
The various $E_\mu$ do not in general commute, and therefore a
detection event does not correspond to what is commonly called the
''measurement of an observable''. Still, the activation of a particular
detector is a macroscopic, objective phenomenon. There is no
uncertainty as to which detector actually clicked. [...]}

\hfill Asher Peres, 2003 \cite[p.1545]{Per2003}
}

\bigskip

\nopagebreak
\hfill\parbox[t]{10.8cm}{\footnotesize

{\em
The only form of ''interpretion'' of a physical theory that I find
legitimate and useful is to delineate approximately the ensemble of
 natural phenomena the theory is supposed to describe and to construct
something resembling a ''structure-preserving map'' from a subset of
mathematical symbols used in the theory that are supposed to represent
physical quantities to concrete physical objects and phenomena
(or events) to be described by the theory. Once these items are
clarified the theory is supposed to provide its own ''interpretation''.
}

\hfill J\"urg Fr\"ohlich, 2019 \cite[p.3]{Fro2019r}
}

\bigskip

To relate q-expectations and q-uncertainties to experimental practice
we employ carefully defined rules, thereby providing a clear formal
notion of measurement.

To measure aspects of a source we introduce a collection of
\bfi{detector elements} labelled by labels $k$ from a finite\footnote{
Experimentally realizable detectors always produce only a finite number
of possible results; see the examples in Section \ref{s.ex}.
Idealizations violating this conditions are not discussed in this paper.
} 
set $K$ satisfying the following postulate.

\bfi{(DRP)}: \bfi{Detector response principle}.
A detector element $k$ responds to a stationary\footnote{
A source is stationary if its properties of interest are
time-independent.
In order that a measured rate has a sensible operational meaning,
the source must be reasonably stationary at least during the time
measurements are taken. Assuming this allows us to ignore all dynamical
issues, including the dynamical differences between isolated systems
and open systems. (In particular, we proceed independently of quantum
mechanical models for the measurement process itself, which involve
microscopical dynamics.) For nonstationary sources, one still gets
time-dependent empirical rates of limited accuracy.
} 
source with density
operator $\rho$ with a rate $p_k$ depending linearly on $\rho$.
Each $p_k$ is positive for at least one density operator $\rho$.

\begin{thm}
If the rates are normalized such that $\sum p_k=1$ then there is a
unique discrete POVM $P_k$ ($k\in K$) such that
\lbeq{e.BornPOVM}
p_k=\Tr\rho P_k =\<P_k\> \for k\in K.
\eeq
\end{thm}

\bepf
For simplicity, we first assume a Hilbert space with finite
dimension $d$; the finiteness restriction is lifted later. By linearity,
the rates satisfy
\lbeq{e.pk}
p_k=\sum_{i,j} P_{kji}\rho_{ij}
\eeq
for suitable complex numbers\footnote{
These numbers can be found operationally by approximately measuring the
rates for at least $d^2$ density operators $\rho$ spanning the space of
$d\times d$ matrices and solving the resulting semidefinite linear least
squares problem for the coefficients. This process is essential for
calibrating detectors and is called \bfi{quantum detection tomography};
see, e.g., \sca{D'Ariano} et al. \cite{DArMP} or \sca{Lundeen} et al.
\cite{Lun}. Of course, to do this one needs sources with known density
operator. In optical applications, textbook quantum optics is used for
these.
} 
$P_{kji}$. If we introduce the matrices $P_k$ with $(j,i)$ entries
$P_{kji}$, \gzit{e.pk} can be written in the concise form
\gzit{e.BornPOVM}.
To find the properties of the matrices  $P_k$ we first note that the
$p_k$ are rates of a stationary process. Hence they are nonnegative and
sum to a constant. Since $p_k>0$ for at least one density operator
$\rho$, \gzit{e.BornPOVM} implies that all $P_k$ are nonzero. Since
$p_k$ is real for all density operators $\rho$, we have
\[
\Tr\rho P_k^*=\Tr(P_k\rho)^*=\ol{\Tr P_k\rho}=\ol p_k=p_k=\Tr\rho P_k.
\]
This holds for all density operators $\rho$, hence $P_k^* = P_k$. Thus
the $P_k$ are Hermitian.
Picking arbitrary pure states with $\rho=\psi\psi^*$ shows that $P_k$
is positive semidefinite. Summing the rates shows that the sum of the
$P_k$ is a multiple of the identity. Requiring this multiple to be $1$
is conventional and amounts to a choice of units for the rates $p_k$
in such a way that they can be interpreted as detection probabilities.
Equivalently, the density operator of the source is normalized to have
trace 1.

Thus, in the case where the Hilbert space of the system measured is
finite-dimensional, the $P_k$ form a discrete POVM. It can be shown
(\sca{de Muynck} \cite[p.41]{deMuy2002}) that the same holds in the
infinite-dimensional case, but the argument is considerably more
abstract. Instead of the matrix argument one uses the fact that each
bounded linear functional on the Hilbert space of Hilbert--Schmidt
operators can be represented as an inner product, resulting in $P_k$s
with $p_k=\Tr\rho P_k$. By the above arguments they are then found to
be positive semidefinite bounded Hermitian operators.
\epf

Formula \gzit{e.BornPOVM}, derived here from very simple first
principles,\footnote{\label{f.3}
The usual practice is to assume Born's rule for projective measurements
as a basic premise. Then the POVM setting is postulated and justified in
terms of Born's rule in an artificial extended  Hilbert space defined
using an appropriate ancilla. This justification is based on Naimark's
theorem (\sca{Naimark} \cite{Nai}) -- also called Neumark's theorem,
using a different transliteration of the Russian originator.
} 
is a well-known extension of von Neumann's formulation of Born's
probability formula. It gives the theoretical q-expectations $\<P_k\>$ a
statistical interpretation in terms of response probabilities of a
quantum detector.

Note that there is a dual result by \sca{Bush} \cite{Bus} that assumes
properties of states in terms of POVMs to prove the existence of a
corresponding density operator $\rho$ satisfying \gzit{e.BornPOVM}.

\subsection{What is measured?}\label{ss.meas}

\nopagebreak
\hfill\parbox[t]{10.8cm}{\footnotesize

{\em
One would naturally like to know what is being measured in a
measurement.}

\hfill Jos Uffink, 1994 \cite[p.205]{Uff1994}
}

\bigskip

\nopagebreak
\hfill\parbox[t]{10.8cm}{\footnotesize

{\em In a statistical description of nature only expectation values
or correlations are observable.
}

\hfill Christof Wetterich, 1997 \cite{Wet}
}
\bigskip

The detection events are usually encoded numerically.
A \bfi{scale} is an assignment of distinct
complex numbers or vectors $a_k$ to the possible detection elements $k$.
In concrete settings, the value assigned by the scale to the $k$th
detection event is whatever has been written on the scale the pointer
registering an event points to, or whatever has been programmed to be
written by an automatic digital recording device.
A \bfi{quantum detector}\footnote{
This may be considered as a technically precise version of the informal
notion of an \bfi{observer} that figures prominently in the foundations
of quantum mechanics. It removes from the latter term all
anthropomorphic connotations.
} 
(in the following simply called a \bfi{detector}) is defined as a finite
collection $K$ of detection elements $k\in K$ of which at most one
responds at any given time, defining a stochastic process of events,
together with a scale $a_k$ ($k\in K$). The POVM part of a detector
description makes no claim about which values are
measured.\footnote{\label{f.POVM}
Here we essentially follow the view of \sca{Schroeck}
\cite{Schr1985,Schr1989}. It is slightly different from the
more traditional point of view as presented, e.g., in \sca{Busch} et al.
\cite{BusGL,BusLPY} or \sca{Holevo} \cite{Hol1982,Hol2001,Hol2012}.
There the POVM is an appropriate family of positive operators
$\Pi(\Delta)$, where $\Delta$ ranges over the subsets of $\Cz^m$ (or
even only of $\Rz$). In the notation of the current setting, their POVM
is given by
\[
\Pi(\Delta):=\sum_{k\in K: a_k\in\Delta} P_k.
\]
Since this implies $\Pi(\{\xi\})=P_k$ if $\xi=a_k$ and
$\Pi(\{\xi\})=\emptyset$ otherwise, the traditional POVM encodes both
the detector elements and the scale, and hence fully specifies the
detector.
\\
Their terminology when restricted to the commutative case (representing
classical physics in the Hilbert space formulation by \sca{Koopman}
\cite{Koo}) amounts to treating all real random variables as observables
(Busch terminology) or generalized observables (Holeovo terminology).
But this is not the view of classical metrology, where 
(cf. \sca{Rabiniwitz} \cite{Rab} and Subsection \ref{ss.measErr} below) 
classical observables always have a true value determined by the 
theoretical description, and all randomness in measurements is assumed 
to be due to noise in the measurement.
} 
It just says that one of the detector
elements making up the detection device responds with a probability
given by the trace formula. The quantum effects are in the response of
the detector elements, not in the scale used to interpret the responses
numerically. The value assigned to the $k$th detection event is a purely
classical convention, and can be any number $a_k$ -- whatever has been
written on the scale the pointer points to, or whatever has been
programmed to be written by an automatic digital recording device.

The results of a detector in a sequence of repeated events define a
random variable or random vector $a_k$ (with a dummy index $k$) that
allows us to define the \bfi{statistical expectation}
\lbeq{e.statEx}
\E(f(a_k)):=\sum_{k\in K} p_kf(a_k)
\eeq
of any function $f(a_k)$. This statistical expectation is operationally
approximated by finite sample means of $f(a)$, where $a$ ranges over
a sequence of actually measured values. However, the statistical
expectation is the usually employed abstraction of this that works with
a probabilistic limit of arbitrarily many measured values, so that the
replacement of relative sample frequencies by probabilities is
justified. Clearly, $\E$ is linear in its argument.
If we introduce for any family $x_k$ ($k\in K$) the operators
\lbeq{e.Px}
P[x_k]:=\sum_{k\in K} x_kP_k,
\eeq
so that in particular,
\[
P[1]=1,
\]
we may use \gzit{e.BornPOVM} to write \gzit{e.statEx} as
\lbeq{e.statExP}
\E(f(a_k))= \<P[f(a_k)]\>.
\eeq
We say that a detector defined by the POVM $P_k$ ($k\in K$) and the
scale $a_k$ ($k\in K$) \bfi{measures} the \bfi{quantity}\footnote{
In traditional terminology (e.g., \sca{Schroeck}
\cite{Schr1985}, \sca{de Muynck} \cite[p.360]{deMuy2002}), one would
say that the detector measures the observable represented by the scalar
or vector operator $A$. Since there is a tradition for using the word
'observable' synonymous with the POVM in the form mentioned in Footnote
${}^{\ref{f.POVM}}$, and since there are lots of observables -- such as
spectral widths or intensities -- that cannot be represented in this
way, we use a more neutral terminology.
} 
\lbeq{e.obs}
A:=P[a_k]=\sum a_kP_k.
\eeq
When the scale consists of real numbers only, the operator corresponding
to the measurement is Hermitian.
From \gzit{e.BornPOVM} and \gzit{e.obs} we find the formula
\lbeq{e.BornEx}
\E(a_k)=\Tr\rho A=\langle A\rangle
\eeq
for the statistical expectation of the measurement results $a_k$
obtained from a source with density operator $\rho$. Comparing with
\gzit{e.qEx}, we see that the statistical expectation of measurement
results coincides with the theoretical q-expectation of $A$ evaluated
in the state $\<\cdot\>$ of the source. This is
\bfi{Born's rule in expectation form}, in the context of
measurements\footnote{
The first published statement of this kind seems to be in the paper by
\sca{Landau} \cite[(4a),(5)]{Landau1927}, but without any reference to
measurement. For a detailed history of the various forms of Born's
rule, see \sca{Neumaier} \cite[Section 3]{Neu.Ifound},
\cite[Chapter 14]{Neu.CQP}.
} 
first stated by \sca{von Neumann} \cite[p.255]{vNeu1927}.
Born's rule gives the purely theoretical notion of a q-expectation a
statistical interpretation in terms of expectations of measurement
results of a quantum detector.

\subsection{Informationally complete POVMs} \label{ss.ic}

\nopagebreak
\hfill\parbox[t]{10.8cm}{\footnotesize

{\em
In this new approach we have a nonuniqueness in places which the old
theory accepted as physically significant. This means that, for
self-adjoint operators with a well-defined physical meaning like spin,
energy, position, momentum, etc., we have now many mathematical formulas
(many nonorthogonal resolultions of the identity). The natural question
arises: what this means, and how to remove such an ambiguity.
Awareness of this nonuniqueness existed in the early papers, but the
question was not worked out.}

\hfill Marian Grabowski, 1989 \cite[p.925]{Grabow}
}

\bigskip

\nopagebreak
\hfill\parbox[t]{10.8cm}{\footnotesize

{\em
This would mean that one can measure all observables of a system in a
single experiment, merely by relabeling the outcomes. This would,
indeed, offer a radical new solution to the joint measurement problem.}

\hfill Jos Uffink, 1994 \cite[p.207]{Uff1994}
}

\bigskip

If the Hilbert space $\Hz$ has finite dimension $d$ and there are
$|K|>d^2$ detector elements, there is a nontrivial relation
\[
\sum_{k\in K} \alpha_kP_k=0
\]
with real coefficients $\alpha_k$ that do not all vanish. Then
\gzit{e.Px} implies $P[a_k+\alpha_k\xi]=P[a_k]=A$ for all $\xi\in \Rz^m$
(or even $\xi\in \Cz^m$). Therefore  the scale is not uniquely
determined by the detector elements and the quantity $A$ measured.

On the other hand, a POVM $P_k$ ($k\in K$) is called
\bfi{informationally complete} if the $P_k$ span the real vector space
$\Pc$ of Hermitian operators with finite trace. This is possible only
when the Hilbert space $\Hz$ has finite dimension $d$ and then requires
$|K|\ge \dim \Pc =d^2$ detector elements. By choosing
the scale appropriately, an informationally complete POVM allows the
measurement of arbitrary vector quantities $A$ since, by definition,
equation \gzit{e.obs} can be solved componentwise for the components of
the $a_k$.
A \bfi{minimal} informationally complete POVM has $|K|=d^2$; then the
POVM and the quantity $A$ measured uniquely determine the scale.

If the Hilbert space has finite dimension $d$, the knowledge of the
probabilities $p_k$ of an informationally complete POVM determines
the associated state and hence its density operator. The density
operator can be found operationally by approximately measuring the rates
and then solving an associated semidefinite linear least squares
problem for $\rho$. This process is essential for calibrating sources
and is called \bfi{quantum state tomography}; see, e.g.,
\sca{Je\v zek} et al. \cite{JezFH}.

\section{Examples}\label{s.ex}

\nopagebreak
\hfill\parbox[t]{10.8cm}{\footnotesize

{\em The very meaning of the concepts involved -- the concept of a
simultaneous measurement of position and momentum, and the concept of
experimental accuracy -- continues to be the subject of discussion.
[...]
Ordinary laboratory practice depends on the assumption that it is
possible to make simultaneous, imperfectly accurate determinations of
the position and momentum of macroscopic objects.}

\hfill Marcus Appleby, 1998 \cite{App1998}
}

\bigskip

Part of the challenge of experimental physics is to devise appropriate
preparation and measurement protocols in such a way that experiments
with desired properties are possible.
Often this is the most difficult aspect of an experiment.
On the other hand, it is also a difficult task to work out
theoretically, i.e., in terms of statistical mechanics rather than
quantum tomography, the right POVM for a given experiment described in
terms of an -- even idealized -- microscopic model; see, e.g.,
\sca{Breuer \& Petruccione} \cite{BreP.QC} and
\sca{Allahverdyan} et al. \cite{AllBN1}.
However, on an informal level, it is easy to give numerous relevant
examples of detectors in the sense defined above.

\subsection{Polarization state measurements}\label{ss.pol}

In continuation of the introductory example we show here how
POVMs model simultaneous measurements of the polarization state $\rho$
of an optical source; cf. \sca{Brandt} \cite{Bra}.
The idea is to split the primary beam emanating from the source by
means of a number of beam splitters into a finite number of
secondary beams labelled by a label set $K$, passing the $k$th secondary
beam through a filter and detecting individual responses at the $k$th
detector element, defined as the screen at which the $k$th secondary
beam ends. If all filters are linear and non-mixing (non-polarizing),
passing the $k$th filter is described by a mapping
$\rho\to T_k\rho T_k^*$, where $T_k$ is a complex $2\times 2$ matrix,
the \bfi{Jones matrix} (\sca{Jones} \cite{Jon}) of the $k$th filter.
Jones matrices for relevant filters include $T=\gamma I$ with
$0<\gamma<1$ (representing simple attenuation) and $T=\phi\phi^*$ with
a normalized vector $\phi\in\Cz^2$ (representing complete polarization
in a complex direction $\phi$). We also introduce a null detector $0$
accounting for the part of the beams absorbed at the filters, assumed
to respond whenever we expect an event but none of the screens shows
one. If we assume that te beam splitters are lossless and that the 
filters are not perfect (so that the Jones matrices are nonsingular), 
it is not difficult to see that this setting defines a detector with 
POVM given by
\[
P_k=c_kT_k^*T_k,~~~P_0=1-\sum_k P_k
\]
for certain constants $c_k>0$ depending on the splitting arrangement.

For multiphoton states, the state space and hence the collection of
possible filters is much bigger, but using linear quantum-optical
networks (\sca{Leonhardt \& Neumaier} \cite{LeoN}) in place of simple
filters one can in essentially the same way design detectors with given
POVMs.

\subsection{Joint measurements of noncommuting quantities}

Joint measurements of position and momentum are often described in
terms of a POVM built from coherent states.
An idealized joint measurement of position and momentum was described
by a coherent state POVM in \sca{Arthurs \& Kelly} \cite{ArtK}, using
infinitely many projectors $|\alpha\rangle\langle\alpha|$ to all
possible coherent states $|\alpha\rangle$, where $\alpha$ is a complex
phase space variable. By discretizing this using a
\bfi{partition of unity} (i.e., a collection of finitely many smooth
nonnegative functions $e_k$ on phase space summing to 1),
these projectors can be grouped into finitely many positive operators
\[
P_k:=\pi^{-1}\int d\alpha e_k(\alpha)|\alpha\rangle\langle\alpha|
\]
corresponding to finite resolution measurements, making it look more
realistic. This would be suitable as a simple analytic example for
presentation in a course.
But to check how accurate an actual joint measurement of position and
momentum fits this construction for some particular partition of unity
would be a matter of quantum tomography!

\subsection{Realistic measurements of position}

In textbooks one often finds idealized hypothetical measurements of
operators with a continuous spectrum, for example position operators.
Realistic position measurements have limited accuracy and range only,
and no sharp boundaries between the position ranges where a particular
detector element responds. This can be modeled by POVMs based on a
partition of unity on configuration space, analogous to the above 
construction for coherent states. Details are given in
\sca{Ali \& Emch} \cite{AliE.meas}.

\subsection{Measuring particle tracks}\label{ss.tracks}

In experimental practice, measurement is often a fairly complex
procedure -- far more complex than the idealized statement of Born's
rule would suggest.
It involves not just reading a pointer but making a model of
the situation at hand, and often involves nontrivial calculations from
raw observations and the model description of the quantities that count
as measurement results.

As a more complicated, concrete example close to experimental practice
we consider a reasonably realistic version of a time projection 
chamber\footnote{
The description of the STAR time projection chamber in
\sca{Anderson} et al. \cite[ Section 5.2]{And} mentions only 2 layers,
so one has to use linear tracks. The LHC uses more layers and a
helical track finder, see \sca{Aggleton} et al. \cite[ Section 5]{Agg}.
} 
(TPC) for the measurement of properties of particle tracks. Here 
obtaining the measurement results requires a significant amount of 
nontrivial computation, not just pointer readings.

In a TPC, what emanates from the source measured passes an arrangement
of wires arranged in $L$ layers of $w$ wires each and generates electric
current signals, ideally exactly one signal per layer. From these
signals, time stamps and positions are computed by a least squares
process (via the Kalman filter), assuming that the track is a helix.
This is the case for a charged particle in a constant magnetic field,
experiencing energy loss in the chamber due to the induced ionizations.
From the classical tracks reconstructed by least squares, the momentum
is computed in a classical way.

The detector can be described by a POVM with an operator for each of
the $w^L$ possible signal patterns. The value assignment is done by a
nontrivial computer program for the least squares analysis and initially
produces a whole particle track.  Part of the information gathered
is discarded; one typically records the computed energy, position and
velocity (or momentum if the mass is known). Momentum and energy are the
quantities of interest for scattering, but for secondary decays one
also needs the decay position. Thus one measures 7-dimensional phase
space vectors (including 3 position coordinates, 3 momentum coordinates,
and the energy) as a very complicated function of the signal patterns.
The POVM operators exist by the general analysis above, though it is
not easy to describe them explicitly in mathematical terms. But this is
not essential for the basic description, which (according to the
introductory quote by Peres) should be given in laboratory terms only.

\subsection{Projective measurements}\label{ss.proj}

\nopagebreak
\hfill\parbox[t]{10.8cm}{\footnotesize

{\em The standard von Neumann description of quantum measurements
applies when all the detector outcomes are well-defined and correspond
to precise classical measurement values that unambiguously reflect the
state of the system. [...]
When measuring a quantum light state, this ideal situation corresponds,
for example, to the case of a perfect photon-number detector
with unit efficiency and no dark counts. Unfortunately, the quantum
measurement description of von Neumann is not valid for measurement
schemes that employ real detectors, which are normally affected by
several imperfections.}

\hfill Zavatta and Bellini, 2012 \cite{ZavB}
}

\bigskip

Complementing the preceding discussion of concrete measurement
arrangements, we now discuss the traditional idealized view of quantum
measurements that goes back to Born, Dirac, and von Neumann.

Rather than postulating Born's rule for projective measurements, as done
in standard textbooks, we derive it here together with all its
ramifications and its domain of validity, from simple, easily motivated
definitions. In particular, the spectral notions appear not as
postulated input as in traditional expositions, but as consequences of
the derivation.

We call a discrete POVM \bfi{projective} if the $P_k$ satisfy the
\bfi{orthogonality relations}
\lbeq{e.orthP}
P_jP_k=\delta_{jk}P_k \for j,k\in K.
\eeq
We say that a detector measuring $A$ performs a
\bfi{projective measurement} of $A$ if its POVM is projective.
Such projective measurements are unstable under imperfection in the
detector. Therefore they are realistic only under special circumstances.
Examples are two detection elements behind
polarization filters perfectly polarizing in two orthogonal directions,
or the arrangement in an ideal Stern--Gerlach experiment for spin
measurement. Most measurements, and in particular all measurements of
quantities with a continuous spectrum, are not projective.

The orthogonality relations imply that $AP_k=a_kP_k$. Since the $P_k$
sum to $1$, any $\psi\in\Hz$ can be decomposed into a sum
$\psi=\sum P_k\psi =\sum\psi_k$ of vectors $\psi_k:=P_k\psi$
satisfying the equation $A\psi_k=AP_k\psi=a_kP_k\psi=a_k\psi_k$.
Therefore $\psi_k$ (if nonzero) is an eigenvector of $A$ (or of each
component of $A$ in case the $a_k$ are not just numbers) corresponding
to the eigenvalue $a_k$ of $A$. Since $P_k^2=P_k=P_k^*$, the $P_k$ are
orthogonal projectors to the eigenspaces of the $a_k$.

When the $a_k$ are numbers,
this implies that $A$ is an operator with a finite spectrum. Moreover,
$A$ and $A^*$ commute, i.e., $A$ is a normal operator, and in case the
$a_k$ are real numbers, a Hermitian, self-adjoint operator. (This is the
setting traditionally assumed from the outset.) When the
$a_k$ are not numbers, our analysis implies that the components of $A$
are mutually commuting normal operators with a finite joint spectrum,
and if all $a_k$ have real components only, the components of $A$ are
Hermitian, self-adjoint operators.
Thus the projective setting is much more limited with respect to the
kind of quantities that it can represent.

For projective measurements, \gzit{e.obs} implies
\[
f(A^*,A)=\sum_k f(\ol a_k,a_k) P_k
\]
for all functions $f$ for which the right hand side is defined.
Therefore the modified scale $f(\ol a,a)$ measures $f(A^*,A)$, as we
are accustomed from classical measurements, and defines a projective
measurement of it. But when the components of $A$ are not normal or do
not commute, this relation does not hold.

From the above discussion we conclude in particular that the possible
values in a projective\footnote{
In a general measurement as discussed in Subsection \ref{ss.meas}, the
measurement results are usually unrelated to the eigenvalues of $A$.
} 
measurement of $A$ are precisely the finitely
many eigenvalues of $A$ (or joint eigenvalues of the components),
measured with a probability of $p_k=\Tr \rho P_k$. This is the textbook
form of \bfi{Born's rule}, valid for projective measurements of
quantities represented by mutually commuting normal operators with a
finite joint spectrum.

In the special case where the spectrum of $A$ is \bfi{nondegenerate},
i.e., all eigenspaces have dimension 1, the orthogonal projectors have
the special form $P_k=\phi_k\phi_k^*$, where $\phi_k$ are normalized
eigenstates corresponding to the eigenvalue $a_k$. In this case, the
probabilities take the form
\[
p_k=\phi_k^*\rho\phi_k.
\]
If, in addition, the source is pure, described by $\rho=\psi\psi^*$ with
the normalized state vector $\psi\in\Hz$, this can be written in the
more familiar squared amplitude form
\lbeq{e.BornSquare}
p_k=|\phi_k^*\psi|^2.
\eeq
In practice, the orthogonality relations \gzit{e.orthP} can be
implemented only approximately (due to problems with efficiency, losses,
inaccurate preparation of directions, etc.). Thus the present
derivation shows that measurements satisfying Born's rule
(i.e., projective measurements) are always idealizations.

Whenever one simultaneously measures quantities
corresponding to noncommuting operators, Born's rule in textbook form
does not apply and one needs a POVM that is not projective.
The operators corresponding to most measurements discussed in Section
\ref{s.ex} do not commute; therefore such joint measurements cannot
even be formulated in the textbook setting of projective measurements.

In general, the POVM description of a real device cannot simply be
postulated to consist of orthogonal projectors. The correct POVM must be
found out by quantum tomography, guided by the theoretical model of the
measuring equipment but ultimately just using the formula
\gzit{e.BornPOVM} for probabilities. This formula is a proper extension
of Born's rule for probabilities of projective measurements.
It cannot be reduced to the latter unless one adds to the description
nonphysical stuff -- namely imagined ancillas without a physical
representation, formally constructed on the basis of Naimark's theorem
(cf. Footnote ${}^{\ref{f.3}}$).

\subsection{A modern introduction to quantum mechanics}
\label{ss.introQM}

The developments so far are suitable for an introductory course on 
quantum mechanics. To introduce POVMs without using the standard
formulation in the usual terms of observables and states is simpler
than to introduce the eigenvalue form of Born's rule in full generality.
To explain the correct rule in an introduction to quantum mechanics is 
easier than writing down Born's rule, because one needs no discussion 
of the spectral theorem and of the subtle problems with self-adjointness
and associated proper boundary conditions. Thus in the foundations, 
there is no longer an incentive for giving a special, highly idealized
case in place of the real thing.

After introducing the basic kinematical framework as in Section 
\ref{s.measP} and illustrating it as in the preceding subsections, the 
next step is to motivate the dynamics. Again, classical optics provides 
the lead.
A linear, non-mixing (not depolarizing) instrument (for example a
polarizer or phase rotator) is characterized by a complex $2\times 2$
\bfi{Jones matrix} $T$. For example, a perfect \bfi{polarizer} has 
the rank 1 form $T=\phi\phi^*$, where $|\phi|^2=1$. 
The instrument transforms an in-going beam with density operator $\rho$
into an out-going beam in the state with density operator 
$\rho'=T\rho T^*$.The intensity of a beam after passing the instrument 
is $S_0'=\Tr \rho'=\Tr T\rho T^*=\Tr \rho T^*T$. 

Passage through an inhomogeneous medium can be modeled by means of many
slices consisting of very thin instruments with Jones matrices close to 
the identity, hence of the form
\lbeq{e.TK}
T(t)=1+\Delta t K(t)+O(\Delta t^2),
\eeq
where $\Delta t$ is the very short time needed to pass through one 
slice and $K(t)$ is an operator specifying how $T(t)$ deviates from the 
identity. If $\rho(t)$ denotes the density operator at time $t$ then
$\rho(t+\Delta t) = T(t)\rho(t)T(t)^*$, so that 
\[
\frac{d}{dt}\rho(t)
= \frac{\rho(t+\Delta t)-\rho(t)}{\Delta t}+O(\Delta t)
= K(t)\rho(t)+\rho(t)K(t)^*+O(\Delta t).
\]
In the continuum limit $\Delta t\to 0$ we obtain the 
\bfi{quantum Liouville equation}
\lbeq{e.qLiou}
\frac{d}{dt}\rho(t) = K(t)\rho(t)+\rho(t)K(t)^*.
\eeq
If the instrument is lossless, the intensities of the in-going and the 
out-going beam are identical. This is the case if and only if the Jones 
matrix $T$ is unitary. Inserting \gzit{e.TK} into the equation $TT^*=1$
and comparing the coefficent of $\Delta t$ shows that $K(t)+K(t)^*=0$. 
Therefore the time-dependent \bfi{Hamiltonian} defined by
\[
H(t)=i\hbar K(t),
\]
is in the lossless case Hermitian, and the quantum Liouville equation
takes the special commutator form 
\lbeq{e.vNeu}
i\hbar\frac{d}{dt}\rho(t) = [H(t),\rho(t)]
\eeq
of the \bfi{von Neumann equation}.

More generally, a linear, mixing (depolarizing) instrument transforms 
$\rho$ instead into a sum of several terms of the form $T\rho T^*$.
It is therefore described by a real $4\times 4$ \bfi{Mueller matrix}
(\sca{Perez \& Ossikovski} \cite{PerO}) acting on the Stokes vector.
Equivalently, it is described by a completely positive linear map 
(\sca{Kossakowski} \cite{Kos}, \sca{Choi} \cite{Choi}) on the space of 
$2\times 2$ matrices, acting on the polarization matrix. Repeating in
this more general situation the derivation of the quantum Liouville 
equation in a more careful manner, taking into account second order 
terms, leads to the \bfi{Lindblad equation} (\sca{Lindblad} \cite{Lin}) 
for the general dynamics of a realistic quantum system.

For the idealized case of beams in a pure state $\psi$, further 
development is possible. For example, the perfect polarizer with 
$T=\phi\phi^*$ reduces the intensity of a pure state $\psi$ to
\[
S_0'=\<T^*T\>=|\phi^*\psi|^2.
\]
This is \bfi{Malus' law} from 1809 (\sca{Malus} \cite{Mal}). 
Reinterpreted in terms 
of detection probabilities, this gives Born's squared amplitude formula 
for quantum probabilities. An instrument with Jones matrix $T$ 
transforms a beam in the pure state $\psi$ into a beam in the pure 
state $\psi'=T\psi$. If $\psi(t)$ denotes the pure state at time $t$ 
then the same slicing scenario as above gives 
$\psi(t+\Delta t) = T(t) \psi(t)$. Therefore 
\[
\frac{d}{dt}\psi(t)
= \frac{\psi(t+\Delta t)-\psi(t)}{\Delta t}+O(\Delta t)
= K(t) \psi(t)+O(\Delta t),
\]
giving in the continuum limit $\Delta t\to 0$ the 
\bfi{time-dependent Schr\"odinger equation}
\[
i\hbar\frac{d}{dt}\psi(t) = H(t) \psi(t).
\]
Thus a polarized quasimonochromatic beam of classical light
behaves exactly like a modern quantum bit. We might say that classical
ray optics in the form known already in 1852 by \sca{Stokes} \cite{Sto} 
is just the quantum physics of a single qubit passing through a medium,
complete with all bells and whistles. This is discussed in some more 
detail in \sca{Neumaier} \cite{Neu.qubit}.

It is now easy to generalize all this to the case of general quantum 
systems. At this point, it makes sense to go through the considerations 
of Sections 2.5 and 2.6 of my book (\sca{Neumaier} \cite{Neu.CQP}) to 
understand the role and limitations of pure states and the  
Schr\"odinger equation for general quantum systems.

Proceding as outlined provides a fully intelligible motivation for all 
basic features of quantum mechanics and quantum information theory.
In contrast to the usual treatments, where the basic features are
addressed by just postulating the required items, usually even in a 
highly idealized form (for pure states and projective measurements), 
this gives actual understanding, not only the appearance of it.

\bigskip

Real systems oscillate and need an infinite-dimensional Hilbert space.
Thus after the qubit, one should introduce the anharmonic oscillator, 
the second simplest system of fundamental importance. This shows that
finite-dimensional Hilbert spaces are not enough and infinitely many
dimensions (i.e., functional analysis) are needed.

Now a lot of elementary phenomena (related to boundary conditions,
bound states and scattering states, tunneling) can be discussed in 
terms of exactly solvable problems. Here the Schr\"odinger equation 
starts to become important, as a computational tool. 

Then one may introduce canonical commutation relations and Ehrenfest's 
theorem for q-expectations. As in Chapter 2 of \sca{Neumaier} 
\cite{Neu.CQP} one may derive the classical limit, where operators may 
be replaced by their q-expectations without introducing significant 
errors. A simple consequence is the \bfi{Rydberg--Ritz formula}
\lbeq{e.RR}
\hbar \omega=E_j-E_k
\eeq
relating a discrete energy spectrum $E_0<E_1<E_2<\ldots$ to observable 
spectral lines with frequencies $\omega$. The observation by 
\sca{Dirac} \cite{Dirac1925} that the Poisson bracket is the 
classical limit of the scaled commutator implies that coupled quantum 
oscillators are described by a tensor product of Hilbert spaces. 
This is easily generalized to arbitrary composite systems. At this 
point, the close connection between classical mechanics and quantum 
mechanics is established.

Now one can introduce annihilation and creation operators for the
harmonic oscillator, and then for a system of $n$ harmonic oscillators.
This motivates bosonic Fock spaces over an $n$-dimensional Hilbert 
space. Proceeding to an infinite number of oscillators, one can turn to 
the interpretation of bosonic Fock space as the Hilbert space of an
arbitrary number of indistinguishable particles.
One can then play with the construction and look at fermionic Fock
spaces. This is the Hilbert space of $n$ qubits, important for quantum
infomation theory.

Then one may raise curiosity about Fock spaces over
infinite-dimensional Hilbert spaces and relate them to quantum fields
and systems of arbitrarily many free particles.

The canonical commutation relations and the Poisson bracket provide 
first examples of the use of Lie algebras in quantum physics. From there
it is only a small step to other important Lie algebras. In particular,
one can restrict the Poisson bracket to rigid bodies and obtains a 
Lie algebra $so(3)$ whose Lie product is the vector product in 3 
dimensions, and whose generators are the components of angular momentum.
It is also the Lie algebra $su(2)$ of the Hermitian quantities with 
zero trace for the qubit, allowing one to present the germs of 
representation theory.

The next step would be to discuss approximation methods and scattering
theory, but this is already beyond the foundation.

\section{Measurement uncertainty}\label{s.unc}

\nopagebreak
\hfill\parbox[t]{10.8cm}{\footnotesize

{\em Some hypotheses are dangerous, first and foremost those which are
tacit and unconscious. And since we make them without knowing them, we
cannot get rid of them. Here again, there is a service that mathematical
physics may render us. By the precision which is its characteristic, we
are compelled to formulate all the hypotheses that we would
unhesitatingly make without its aid.}

\hfill Henri Poincar\'e, 1902 \cite[p.151]{PoiScH}
}

\bigskip

This section gives a thorough, precise discussion of various aspects of 
uncertainty in quantum measurements. There are natural links to the 
thermal interpretation of quantum physics as defined in my book
(\sca{Neumaier} \cite{Neu.CQP}).
Subsections \ref{ss.rep}--\ref{ss.measTI} use material from Sections 
10.6-10.7 of this book.

\subsection{Statistical uncertainty}\label{ss.unc}

\nopagebreak
\hfill\parbox[t]{10.8cm}{\footnotesize

{\em Aus diesen Gr\"unden ist eine gleichzeitige genaue Beobachtung
von $q$ und $p$ prinzipiell ausgeschlossen. [...]
Man kann aber auch beide Gr\"o{\ss}en in einer einzigen Beobachtung
messen, also wohl gleichzeitig, aber nur mit beschr\"ankter Genauigkeit.
Bei einer solchen Beobachtung fragt man in der klassischen Theorie nach
dem 'Fehler' des gemessenen Wertes. [...]
Die 'Beobachtungsfehler' erscheinen in der neuen Theorie als
mit der statistischen Unbestimmtheit selbst zusammengeschmolzen.}

\hfill Earle Kennard 1927 \cite[p.340f]{Ken}
}

\bigskip

\nopagebreak
\hfill\parbox[t]{10.8cm}{\footnotesize

{\em
Results of measurements cannot be absolutely accurate. This
unavoidable imperfection of measurements is expressed in their
inaccuracy.}

\hfill Semyon Rabinovich, 2005 (\cite[p.2]{Rab})
}

\bigskip

We write $|x|:=\sqrt{x^*x}$ for the Euclidean norm of a vector
$x\in \Cz^m$, and generalize it to vectors $A\in(\Lin\Hz)^m$  with
operator components by defining\footnote{
Actually we never need $|A|$ but only the notation $|A|^2$ for $A^*A$,
especially when $A$ is some composite formula. 
} 
the operator
\[
|A|:=\sqrt{A^*A}.
\]
This allows us to formulate and prove the basic inequality
\lbeq{e.Esigma}
\min_{\xi\in\Cz^m}\E(|a_k-\xi|^2)=\E(|a_k-\ol A|^2)
\ge \<|A-\ol A|^2\>=\sigma_A^2
\eeq
bounding the statistical uncertainty of the measurement results $a_k$
in terms of the theoretical q-uncertainty of the quantity $A$ measured.
This result, for $m=1$ due to \sca{Holeveo} \cite[(9.8), p.88]{Hol1982}
and later redicovered by \sca{de Muynck \& Koelman} \cite{deMuyK} and
\sca{Werner} \cite[Proposition 3(2)]{Wer}, follows by observing that
\lbeq{e.Eaxi}
\E(|a_k-\xi|^2)-\E(|a_k-\ol A|^2)=\E(|a_k-\xi|^2-|a_k-\ol A|^2)
=\E(|\ol A-\xi|^2)=|\ol A-\xi|^2
\eeq
is minimal for $\xi=\ol A$. Now the following proposition (cf.
\sca{Holevo} \cite[Lemma 13.1]{Hol1973} for the case $m=1$) applies.

\begin{prop}\label{p.Esigma}
Given a detector measuring $A=P[a_k]\in\Hz^m$. Then:

(i) For every Hermitian positive semidefinite $G\in\Cz^{m\times m}$, the
operator $P[a_k^*Ga_k]-A^*GA$ is positive semidefinite.

(ii) For any $\xi\in\Cz^m$ and any state $\<\cdot\>$,
\lbeq{e.Exi}
\E(|a_k-\xi|^2)\ge \<|A-\xi|^2\>.
\eeq
\end{prop}

\bepf
Using the POVM $P_k$ ($k\in K$) of the detector, we write
\[
\Delta:=\sum_{k\in K} (A-a_k)^*GP_k(A-a_k).
\]
(i) For any $\psi\in\Hz$, we define the vectors $\psi_k:=(A-a_k)\psi$
and find $\psi^*N\psi=\sum \psi_k^* GP_k\psi_k\ge 0$. Hence $\Delta:$ is
positive semidefinite. Since
\[
\bary{lll}
\Delta:&=&\D\sum\Big(A^*GP_kA-a_k^*GP_kA-A^*GP_ka_k+a_k^*GP_ka_k\Big)\\
&=&A^*GP[1]A-A^*GP[a_k]-P[a_k]^*GA+P[a_k^*Ga_k]\\
&=&A^*GA-A^*GA-A^*GA+P[a_k^*Ga_k]=P[a_k^*Ga_k]-A^*GA,
\eary
\]
the first part follows.

(ii) In the special case where $G$ is the identity matrix, we find from
\gzit{e.statExP} that
\[
\E(|a_k|^2)=\<P[|a_k|^2]\>=\<P[a_k^*a_k]\>=\<N+A^*A\>
=\<\Delta:\>+\<A^*A\>\ge \<A^*A\>=\<|A|^2\>.
\]
Since $P[a_k-\xi]=P[a_k]-\xi=A-\xi$ we may apply this with
$a_k-\xi$ in place of $a_k$ and $A-\xi$ in place of $A$ and find
\gzit{e.Exi}.
\epf

In the special case of projective measurements, $A=P[a_k]$ satisfies
\[
P_k(A-a_k)=P_kA-a_kP_k=\sum_j a_jP_kP_j-a_kP_k=0,
\]
so that $\Delta:=0$ in the proof of Proposition \gzit{p.Esigma}.
Therefore inequality \gzit{e.Esigma} holds in this case with equality 
for all states. A converse was proved in \sca{Kruszy\'nski \& De Muynck}
\cite[Proposition 2]{KruM}. Thus the difference in \gzit{e.Esigma}
describes the lack of projectivity. We may view it as a measure of
quality of a real detector, as opposed to an idealized projective one.
\sca{Busch} et al. \cite[(16)]{BusHL} view this difference as a measure
of intrinsic noise.

\subsection{Imperfect measurements}\label{ss.imp}

\nopagebreak
\hfill\parbox[t]{10.8cm}{\footnotesize

{\em
The fact that actual measurements are always imprecise is well-known
and led Poincare to distinguish carefully the "mathematical continuum"
from the "physical continuum." In the mathematical continuum the notion
of identity satisfies the usual transitivity condition [...]
By contrast, this property cannot be assumed for the notion of
"indistinguishability" in the physical continuum attached to the raw
data of experiments. [...]
To pass from the physical continuum to the mathematical continuum
requires an idealization, namely that infinitely precise measurements
are in principle, if not in fact, attainable.}

\hfill Ali and Emch, 1974 \cite[p.1545]{AliE.meas}
}

\bigskip

\nopagebreak
\hfill\parbox[t]{10.8cm}{\footnotesize

{\em The discrete nature of the reading scale entails that a given
measuring apparatus allows only a measurement of a discrete version of
the observable under consideration. With this we do not, however, deny
the operational relevance of continuous observables. On the contrary,
their usefulness as idealisations shows itself in the fact that they
represent the possibility of indefinitely increasing the accuracy of
measurements by choosing increasingly refined reading scales.
}

\hfill Busch, Lahti and Mittelstaedt, 1996 (\cite[p.81]{BusLM})
}

\bigskip

The spectral norm of a quantity measured by an arbitrary detector is
bounded since the sum \gzit{e.obs} is finite and $\|P_k\|\le 1$ for all
$k$. Hence the components of $A$ are bounded linear
operators.\footnote{
To extend the notion of measurement to unbounded quantities such as
position or momentum -- which is outside the scope of the present paper
--, one would need to proceed in an idealized fashion, using continuous
POVMs for idealized measurements with infinite precision. Then
q-expectations are defined only for sufficiently regular density
operators. For a proper treatment of the unbounded case see, e.g., the 
books mentioned at the beginning of this paper.
} 
In particular, \gzit{e.BornEx} is defined for all
density operators $\rho$. Note that the same bounded linear operator
$A$ can be decomposed in many ways into a linear combination of the
form \gzit{e.obs}; thus there may be many different detectors with
different scales measuring quantities corresponding to the same
operator $A$.

In practice, one is often interested in measuring a given (bounded or
unbounded) operator $X$ of interest. Designing realistic detectors that
allow a high quality measurement corresponding to theoretically
important operators is the challenge of high precision experimental
physics. Due to experimental limitations, this generally involves both
statistical and systematic errors.

If we approximate a (possibly vector-valued) quantity $X$ by a
measurable substitute quantity $A$ -- some such approximation is
unavoidable in practice --, we make a systematic error that may depend 
on the state of the system measured. \gzit{e.Eaxi} implies the formula
\lbeq{e.Emse}
\E(|a_k-\ol X|^2)=\E(|a_k-\ol A|^2)+|\ol A-\ol X|^2
\eeq
for the mean squared error of $a_k$ as an approximation of $\ol X$.
In particular, the term
\[
\Delta:=|\ol A-\ol X|=|\<X\>-\<A\>|,
\]
the \bfi{bias} due to the substitution of $A$ for $X$, is a lower bound
for the \bfi{root mean squared error} (\bfi{RMSE})
\[
\eps_X:=\sqrt{\E(|a_k-\ol X|^2)}.
\]
Unlike the RMSE but like q-expectations, the bias is a theoretical
property of a state, independent of measurement.
We say that $A$ is an \bfi{unbiased} approximation of $X$ in all states
such that the bias vanishes. If the bias vanishes in all states from an
open subset of the set of all states, we necessarily have $X=A$.
In particular, there are no everywhere unbiased approximations of
unbounded operators.\footnote{
For unbounded operators $A$ (for example the position operator vector
$q$ in a particular coordinate system), formula \gzit{e.BornEx} and
hence the bias is well-defined only for sufficiently regular density
operators $\rho$. This reflects a slight deficiency of the standard
textbook presentation of expectations. However,
measurement equipment for unbounded operators such as position, say,
only produces results in a bounded range. Hence it corresponds in fact
to a measurement device for a clipped version $X=F(q)$ of the
position operator $q$ with bounded $F$, resulting in a bounded $X$.
} 
Note that in practice, $\ol A \ne \ol X$ due to imperfections.
In particular, unbiased measurements are necessarily idealizations.

The RMSE $\eps_X$ measures the uncertainty in the value assigned
to $\ol X$. But because of the broken watch effect,\footnote{
Measuring time with a broken watch shows twice a day the exact time,
whereas a watch that is slow 1 second per day shows the correct time
at most once in a century.
} 
it cannot be regarded as the measured uncertainty in the value assigned
to $X$. Thus we need to add a systematic uncertainty correction that
corrects for the possibility that the uncertainty of $A$ is less than
the uncertainty of $X$. The latter has nothing to do with measurement
and hence must be a theoretical quantity computable from
q-uncertainties. Now
\lbeq{e.qmse}
\<|A-\ol X|^2\>=\<|A-\ol A|^2\>+|\ol A-\ol X|^2
=\sigma_A^2+|\ol A-\ol X|^2
\eeq
in analogy to \gzit{e.Emse}, proved by expanding all squares. Comparing
this with the formula
\[
\<|X-\ol A|^2\>=\<|X-\ol X|^2\>+|\ol X-\ol A|^2
=\sigma_X^2+|\ol X-\ol A|^2
\]
obtained by interchanging the role of $A$ and $X$, we see that
the natural uncertainty correction to the mean squared error is
$(\sigma_X^2-\sigma_A^2)_+$, where $x_+:=\max(x,0)$ denotes the positive
part of a real number $x$. We therefore regard
\lbeq{e.measError}
\Delta_X[a_k]:=\sqrt{\E(|a_k-\ol X|^2)+(\sigma_X^2-\sigma_A^2)_+},
\eeq
the square root of the corrected mean squared error,
as the \bfi{measurement uncertainty} when measurements of $A$ are
performed in place of measurements of $X$.
Such \bfi{imperfect measurements} make sense even for unbounded
quantities $X$ in states with finite q-uncertainty $\sigma_X$.
By \gzit{e.Exi} and \gzit{e.qmse},
\[
\bary{lll}
\Delta_X[a_k]^2&=&\E(|a_k-\ol X|^2)+(\sigma_X^2-\sigma_A^2)_+\\
&\ge& \<|A-\ol X|^2\>+\sigma_X^2-\sigma_A^2\\
&=&\sigma_X^2+|\ol A-\ol X|^2
\eary
\]
hence the measurement uncertainty is bounded from below by theoretical
error measures,
\[
\Delta_X(a_k)\ge \sqrt{\sigma_X^2+|\ol A-\ol X|^2}
=\sqrt{\<|X-\ol A|^2\>}.
\]
In particular, $\Delta_X(a_k)$ is always at least as large as the
q-uncertainty of $X$, and larger if there is a nonzero bias.
This gives an operational interpretation to these theoretical terms.

By changing the scale of a detector we may define measurements of many
different quantities $A$ based on the same POVM. By picking the scale
carefully one can in many cases choose it such that $A$ approximates a
particular operator $X$ of interest with small measurement uncertainty
$\Delta_X[a_k]$ for the collection of states of interest. There are
several alternative ways to quantify what constitutes adequate
approximations; see, e.g., \sca{Appleby} \cite{App1998,App2016},
\sca{Barcielli} et al. \cite{BarGT},
\sca{Busch} et al. \cite{BusHL,BusLPY},
and references there.

Finding a good match of $A$ and $X$ by choosing a good scale $a_k$ is
the process called \bfi{tuning}. It corresponds to the classical
situation of labeling the scale of a meter to optimally match a desired
quantity. If the detector can also be tuned by adjusting parameters
$\theta$ affecting its responses, the operators $P_k=P_k(\theta)$
depend on these parameters, giving
\[
A=A(\theta)=\sum a_kP_k(\theta).
\]
Now both the labels $a_k$ and the parameters $\theta$ can be tuned to
improve the accuracy with which the desired $X$ is approximated by
$A(\theta)$, perfecting the tuning.

\subsection{Reproducibility}\label{ss.rep}

\nopagebreak
\hfill\parbox[t]{10.8cm}{\footnotesize

{\em A student has read such and such a number on his thermometer.
He has taken no precautions. It does not matter; he has read it, and if
it is only the fact which counts, this is a reality [...]
Experiment only gives us a certain number of isolated points. They must
be connected by a continuous line, and this is a true generalisation.
But more is done. The curve thus traced will pass between and near the
points observed; it will not pass through the points themselves.
Thus we are not restricted to generalising our experiment, we correct
it; and the physicist who would abstain from these corrections, and
really content himself with experiment pure and simple, would be
compelled to enunciate very extraordinary laws indeed.}

\hfill Henri Poincar\'e, 1902 \cite[p.142f]{PoiScH}
}

\bigskip

\nopagebreak
\hfill\parbox[t]{10.8cm}{\footnotesize

{\em The purpose of measurements is the determination of properties of
the physical system under investigation.
}

\hfill Busch, Lahti and Mittelstaedt, 1996 (\cite[p.25]{BusLM})
}

\bigskip

\nopagebreak
\hfill\parbox[t]{10.8cm}{\footnotesize

{\em
Let us emphasize again that when a qubit is measured, it only ever
gives 0 or 1 as the measurement result -- probabilistically.}

\hfill Nielsen and Chuang, 2001 (\cite[p.14]{NieC})
}

\bigskip

\nopagebreak
\hfill\parbox[t]{10.8cm}{\footnotesize

{\em Measurements are regarded metrologically to be better the lower
their uncertainty is. However, measurements must be reproducible,
because otherwise they lose their objective character and therefore
become meaningless.}

\hfill Semyon Rabinovich, 2005 (\cite[p.22]{Rab})
}

\bigskip

\nopagebreak
\hfill\parbox[t]{10.8cm}{\footnotesize

{\em In a fundamentally statistical theory like quantum mechanics the
results of individual measurements tell us almost nothing: It is always
the probability distribution of outcomes for a fixed experimental
arrangement which can properly be called the result of an experiment.}

\hfill Busch, Lahti and Werner 2014 \cite[p.5]{BusLW}
}

\bigskip

\nopagebreak
\hfill\parbox[t]{10.8cm}{\footnotesize

{\em The Born measure is a mathematical construction; what is its 
relationship to experiment? This relationship must be the source of the 
(alleged) randomness of quantum mechanics, for the Schr\"odinger 
equation is deterministic.}

\hfill Klaas Landsman, 2019 \cite[p.20]{Lan2019} 
}

\bigskip

Let us consider the measurement of a quantity $A\in\Cz^{2\times 2}$ of
an arbitrary 2-state system (a qubit).
According to the experimental record, the response of a sufficiently
good detector produces measurement results concentrated near two spots
(or parallel lines) of the detector, just as what
one gets when measuring a classical diffusion process in a double-well
potential (see, e.g., \sca{Hongler \& Zheng} \cite{HonZ}).
For example, this happens in the original Stern-Gerlach experiment; cf.
the quote of \sca{Busch} et al. \cite{BusGL} at the beginning of
Section \ref{s.measP}. This results in a bimodal distribution
with two more or less sharp peaks, with details depending on the
detection method used and its resolution.

In a frequently used idealization -- e.g., in the typical textbook
treatment of a spin measurement -- one ignores the limited efficiency
of a detector. Then the distribution may even be assumed to be 2-valued,
with measurement results that take only one of two values $\lambda_{1}'$
and $\lambda_{2}'$, corresponding to the two modes of the bimodal
distribution.

In the standard formulation of \bfi{Born's statistical interpretation}
of quantum mechanics, based on projective measurements, the measurement
results are \bfi{quantized}: the measured result will be one of the
eigenvalues $\lambda_k$ of $A$.
Multiple repetition of the measurement results in a random sequence of
values $\lambda_k$, with probabilities computed from \gzit{e.BornSquare}
if the system is in a pure state. In the limit of arbitrarily many
repetitions, the mean value of this sequence approaches $\ol A$ and the
standard deviation approaches $\sigma_A$.

Returning to the qubit case, we assume that $A$ has unknown but distinct
eigenvalues $\lambda_1,\lambda_2$. The q-expectation and the
q-uncertainty of $A$ can be exactly calculated in terms of the
probability $p=p_1$. We assume for simplicity that the system is in a
pure state $\alpha_1|\lambda_1\>+\alpha_2|\lambda_2\>$, where the kets
denote the eigenstates of $A$ and $|\alpha_1|^2=p$, $|\alpha_2|^2=1-p$.
Then q-expectation and q-uncertainty are found to be 
\[
\ol A=p\lambda_1+(1-p)\lambda_2,~~~
\sigma_A=|\lambda_1-\lambda_2|\sqrt{p(1-p)}.
\]
The prediction made by Born's rule is that the observed bimodal 
distribution has point support at the nodes $\lambda_{1}'=\lambda_{1}$
and $\lambda_{2}'=\lambda_{2}$. Clearly, Born's rule only describes
idealized\footnote{\label{f.irrat}
Originally, Born's statistical interpretation was stated
only for energy measurements for systems with discrete energy
levels. Thus the measured quantity is the Hamiltonian $H$,
and its eigenvalues are in general irrational. A measurement according
to Born's rule in its standard form would produce these irrational
numbers exactly. This is clearly not the case. Thus one is forced to 
use a more liberal reading of Born's rule, where some additional
measurement error is acceptable. This means that Born's rule is no 
longer about real measurements but about idealized measurements whose 
observations are theoretical numbers, not actual results. Therefore, in 
this liberal reading, Born's rule is a purely theoretical construct, 
silent about actual measurement results.
} 
measurement values, the eigenvalues $\lambda_k$ of $A$.
The deviations from the eigenvalues, e.g. in realistic Stern--Gerlach
experiments, must be interpreted as measurement errors.

As an illustration we
consider some piece of digital equipment with 3 digit display measuring
some physical quantity $A$ using $N$ independent measurements. Suppose
the measurement results were 6.57 in 20\% of the cases and 6.58 in 80\%
of the cases. Every engineer or physicist would compute the mean
$\ol A= 6.578$, the variance 
\[
\sigma_X^2=\<X^2\>-\<X\>^2=0.2\cdot 0.008^2+0.8\cdot 0.002^2=0.004^2, 
\]
and the standard deviation
$\sigma_A=0.004$, concluding that the true value of the quantity $A$
deviates from $6.578$ by an error of the order of $0.004N^{-1/2}$.
Note that, as always when one measures something with a digital device,
the error distribution is discrete, not Gaussian.

Now we consider the measurement of the Hermitian quantity
$A\in\Cz^{2\times 2}$ of a 2-state quantum system in the pure up state,
using $N$ independent measurements, and suppose that we obtain exactly
the same results. Now Born's statistical interpretation
proceeds differently and claims that there is no measurement error.
Instead, each measurement result reveals one of the eigenvalues
$x_1=6.57$ or $x_2=6.58$ in an unpredictable fashion with probabilities
$p=0.2$ and $1-p=0.8$, up to statistical errors of order $O(N^{-1//2})$.
In particular, the measurement results are not reproducible; only their
statistic is.

For $A=\pmatrix{6.578 & 0.004 \cr 0.004 & 6.572}$, both the engineering
view and Born's interpretation of the results for the 2-state quantum
system are consistent with the data. However, Born's statistical
interpretation deviates radically from engineering practice, without
any apparent necessity. It does not even conform to the notion of
a measurement in the traditional sense since an essential element in 
the latter's specification -- the reproducibility of the result -- is 
not guaranteed. Shouldn't we rather proceed as before and draw the same 
conclusions as the engineer?

According to our discussion in Section \ref{s.measP} in a realistic 
measurement, the possible values obtained when measuring a particular
quantity $A$ depend on the decomposition $A=P[a_k]$ used to construct
the scale. That this decomposition is ambiguous follows from Subsection
\ref{ss.ic}, where we saw that the scale is not determined by the
quantity $A$ measured. Since the scale determines the measurement
results, this means that one can with equal right ascribe different
results to the measurement of the same quantity $A$. Thus, in general,
different detectors measuring the same quantity $A$ have different sets
of possible measurement results.
In particular, the approach introduced in this paper gives projective
measurements (and hence eigenvalues) no longer a special status. The 
same quantity $A$ can be measured by detectors with different 
mathematical characteristics and in particular different measurement 
results that generally have nothing to do with the eigenvalues of $A$.

We conclude that not the individual observations but only their 
statistical properties -- POVM probabilities and expectation values --
are reproducible, hence the latter (and only these) correspond to 
objective properties of the source measured. Thus realistic rules for 
measurement make the eigenvalue link of the traditional interpretation 
of quantum mechanics look artificial.

\subsection{Measurement errors}\label{ss.measErr}

\nopagebreak
\hfill\parbox[t]{10.8cm}{\footnotesize

{\em
Von dem neuen Standpunkt
schwebt nun zun\"achst dieser Fehlerbegriff in der Luft, denn er setzt
doch nicht nur den Begriff elnes beobachteten, sondern auch den
Begriff eines 'wahren' Wertes voraus, und letzteren gibt es im
physikalischen Sinne nicht mehr.}

\hfill Earle Kennard 1927 \cite[p.340]{Ken}
}

\bigskip

\nopagebreak
\hfill\parbox[t]{10.8cm}{\footnotesize

{\em This mean value $x_0$ locates the wave packet in the crude sense
that an observing apparatus must be placed near $x_0$ if it is to have
a significant chance of interacting with the particle.}

\hfill Carl Helstrom, 1974 \cite[p.454]{Hel74}
}

\bigskip

\nopagebreak
\hfill\parbox[t]{10.8cm}{\footnotesize

{\em Measurement is the process of determinating the value of a physical
quantity experimentally with the help of special technical means called
measuring instruments. The value of a physical quantity [...] is found 
as the result of a measurement.
The true value of a measurand is the value of the measured physical
quantity, which, being known, would ideally reflect, both qualitatively
and quantitatively, the corresponding property of the object.}

\hfill Semyon Rabinovich, 2005 (\cite[p.1f]{Rab})
}

\bigskip

\nopagebreak
\hfill\parbox[t]{10.8cm}{\footnotesize

{\em Measurement errors are in principle unavoidable, because a
measurement is an experimental procedure and the true value of the
measurable quantity is an abstract concept. }

\hfill Semyon Rabinovich, 2005 (\cite[p.11]{Rab})
}

\bigskip

Measurement errors are ubiquitous in physical practice; their
definition requires, however, some care. A single measurement produces
a number, the \bfi{measurement result}. The splitting of the measurement
result into the sum of an intended result -- the \bfi{true value} -- and
a \bfi{measurement error} (the deviation from it) depends on what one
declares to be the true value. Thus what can be said about measurement
errors depends on what one regards as the true value of something
measured.

In general, the true value is necessarily a theoretical construct, an
idealization arrived at by convention.
Since measured are only actual results, never the hypothesized true
values, there is no way to determine experimentally which convention is
the right one. Both the quantum formalism and the experimental record
are independent of what one declares to be the true value of a
measurement. Different conventions only define different ways of
bookkeeping, i.e., different ways of splitting the same actual
measurement results into a sum of true values and errors, in the
communication about quantum predictions and experiments. Nothing in the
bookkeeping changes the predictions and the level of their agreement
with experiment.

Thus the convention specifying what to consider as true values is
entirely a matter of choice, an \bfi{interpretation}. The convention
one chooses determines what one ends up with, and each interpretation
must be judged in terms of its implications for convenience and
accuracy. Like conventions about defining measurement units
\cite{SIunits}, interpretations can be adjusted to improvements in
theoretical and experimental understanding, in order to better serve
the scientific community.

According to Born's statistical interpretation in the standard
formulation (i.e., for projective measurements),
each actual measurement result $\lambda$ is claimed to be one of the
eigenvalues, which is exactly (according to the literal
reading\footnote{\label{f.proj}
The formulation appearing in \sca{Wikipedia} \cite{Wik.Born} is
''the measured result will be one of the eigenvalues''.
\sca{Griffiths \& Schroeter} \cite[p.133]{GriS} declare, ''If you
measure an observable [...] you are certain to get one of the
eigenvalues''. \sca{Peres} \cite[p.95]{Peres} defines, ''each one of
these outcomes corresponds to one of the eigenvalues of $A$; that
eigenvalue is then said to be the result of a measurement of $A$''.
The only exceptions seem to be textbooks such as
\sca{Nielsen \& Chuang} \cite[p.84f]{NieC} that start the formal
exposition with the POVM approach rather than Born's interpretation.
But in their informal introduction of qubits, even they give priority 
to projective measurements!
} 
 of most formulations) or approximately (in a more liberal
reading) measured, with probabilities computed from $A$ and the density
operator $\rho$ by the probability form of Born's rule. Since we saw
that deviations from the eigenvalues in realistic experiments must
be interpreted as measurement errors, we conclude that the eigenvalues
are the true values of Born's statistical interpretation.

In the preceding subsections \ref{ss.unc} and \ref{ss.imp}, we did not
assume a notion of true value in the quantum case.
However, \gzit{e.Esigma} implies that the least uncertain value $\xi$
is the q-expectation $\ol A$. Thus in a statistical sense, the best
possible value that can be assigned is $\ol A$. This suggests that in
the unbiased case, $\ol A$ should be (in analogy to classical
statistics) the true value of $A$, measured approximately.

\subsection{The thermal interpretation of quantum physics}
\label{ss.thermal}

\nopagebreak
\hfill\parbox[t]{10.8cm}{\footnotesize

{\em We assume that (i) the density matrix is observable and
(ii) any observable is a function of the density components.}

\hfill Lajos Diosi, 1988 \cite[p.2887]{Dio}
}

\bigskip

\nopagebreak
\hfill\parbox[t]{10.8cm}{\footnotesize

{\em The idea of unsharp objectification arises if one intends to leave
quantum mechanics intact and still tries to maintain a notion of real
and objective properties.
}

\hfill Busch, Lahti and Mittelstaedt, 1996 (\cite[p.127]{BusLM})
}

\bigskip

\nopagebreak
\hfill\parbox[t]{10.8cm}{\footnotesize

{\em 
We shall study a simple measurement problem -- the measurement of the
diameter of a disk. [...] It may happen that the difference of the
measurements in different directions exceeds the permissible error of a
given measurement. In this situation, we must state that within the
required measurement accuracy, our disk does not have a unique diameter,
as does a circle. Therefore, no concrete number can be taken, with
prescribed accuracy, as an estimate of the true value of the measurable
quantity.}

\hfill Semyon Rabinovich, 2005 (\cite[p.11]{Rab})
}

\bigskip

The preceding analysis suggests that we should perhaps reject the 
convention that declares the eigenvalues of operators to be the true
values in a measurement. This is done explicitly in the
\bfi{thermal interpretation of quantum physics} introduced in
\sca{Neumaier} \cite{Neu.IIfound,Neu.IIIfound,Neu.IVfound}; a detailed,
definitive account is in my recent book {\it Coherent quantum physics}
(\sca{Neumaier} \cite{Neu.CQP}); see especially Section 9.2. The thermal
interpretation gives a new foundational perspective on quantum mechanics
and suggests different questions and approaches than the traditional
interpretations.

The thermal interpretation generalizes the well-known fact that in
equilibrium statistical thermodynamics, all extensive quantities are
represented by q-expectations. It proclaims
-- in direct opposition to the tradition created in 1927 by Jordan,
Dirac, and von Neumann -- the alternative convention that the true
values are the q-expectations rather than the eigenvalues. 
In the thermal interpretation of quantum physics, given a particular 
instance of a quantum system described by a model at a given time, the 
state defines all its properties, and hence what \bfi{exists}\footnote{
This gives a clear formal meaning to the notion of existence. Whether
something that exists in this model sense also exists in Nature depends
on how faithful the model is to the corresponding aspect of Nature.
} 
in the system at that time. The objective \bfi{properties} of the system
are given by q-expectations and what is computable from these. 
All properties of a quantum system at a fixed time depend on the state 
$\<\cdot\>$ of the system at this time and are expressed in terms of 
definite but uncertain values of the quantities. As discussed in detail 
in my book (\sca{Neumaier} \cite{Neu.CQP}), the identification of 
these formal properties with real life properties is done by means of

\bfi{(CC)} \bfi{Callen's criterion} (cf. \sca{Callen} \cite[p.15]{Cal}):
{\it Operationally, a system is in a given state if its properties are
consistently described by the theory for this state.
} 

This is a concise version of the principle of identification suggested
by \sca{Eddington} \cite[p.222]{Edd}; cf. the initial quote of this 
paper. Callen's criterion is enough to find out in each single case how 
to approximately determine the uncertain value of a quantity of 
interest. 

As we have seen, the simplest quantum system, a qubit, was already 
described by \sca{Stokes} \cite{Sto} in 1852, in terms essentially 
equivalent to the thermal interpretation -- except that the intrinsic 
uncertainty was not yet an issue, being at that time far below the 
experimentally realizable accuracy. This alternative convention also 
matches the actual practice in quantum information theory, where the 
states are manipulated, transmitted, measured, hence their properties 
(i.e., whatever is computable from the state) are 
treated as objective properties. The density operator may be viewed 
simply as a calculational tool for obtaining these objective properties,
in particular q-expectations and their uncertainties.

In the thermal interpretation, every observable scalar or vector 
quantity $X$ has an associated intrinsic state-dependent uncertainty 
$\sigma_X$ within which it can be (in principle) determined. The idea 
is that the q-expectation $\ol X$ itself has no direct operational 
meaning; only the fuzzy region of $\xi\in\Cz^m$ with $|\xi-\ol X|$ 
bounded by the uncertainty $\sigma_X$ or another small multiple of 
$\sigma_X$ is meaningful. Statistics enters whenever a single value 
has too much uncertainty, and only then. In this case, the uncertainty 
can be reduced -- as within classical physics -- by calculating 
statistical means. 

This is standard engineering practice when considering the diameter of a
disk that is not perfectly circular. The uncertainty is in the imprecise
definition, just as that in the position of an extended object such as 
a doughnut. In particular, the description of a quantum particle as 
having momentum $p$ and being at position $q$ is as unsharp as the 
description of a classical signal as having frequency $\nu$ at time $t$.
Even formally, the concepts are
analogous and share the uncertainty relation, known in signal analysis
as the Nyquist theorem, and discovered in the quantum context by 
Heisenberg. The analogy is especially clear in quantum field
theory, where on the one hand position and time and on the other hand 
momentum and energy, related to frequency by the Rydberg--Ritz formula 
\gzit{e.RR}, are described on an equal footing.
When measuring $X$ it is meaningless to ask for more accuracy than the 
uncertainty $\sigma_X$, just as meaningless as to ask for the position
of a doughnut to mm accuracy. 

Thus the foundations invoked by the thermal interpretation are 
essentially the foundations used everywhere for uncertainty 
quantification, just slightly extended to accommodate quantum effects 
by not requiring that quantities commute.

\subsection{Measurement in the thermal interpretation}
\label{ss.measTI}

\nopagebreak
\hfill\parbox[t]{10.8cm}{\footnotesize

{\em
A satisfactory theory of the measuring process must start from a
characterization of the macroscopic properties of a large body. Such
properties must have an objective character.}

\hfill Daneri, Loinger and Prosperi, 1962 \cite[p.305]{DanLP}
}

\bigskip

\nopagebreak
\hfill\parbox[t]{10.8cm}{\footnotesize

{\em Quantum mechanics has often been classified as a merely
statistical ensemble theory, with not much bearings on the individual
members of the ensembles. Yet there is an increasing variety of
experiments exhibiting individual quantum processes which were
conceived, devised and explained on the basis of this very theory.
}

\hfill Busch and Lahti, 1996 \cite[p.5899]{BusL}
}

\bigskip

\nopagebreak
\hfill\parbox[t]{10.8cm}{\footnotesize

{\em Decoherence actually aggravates the measurement problem: where 
previously this problem was believed to be man-made and relevant only 
to rather unusual laboratory situations, it has now become clear that 
''measurement'' of a quantum system by the environment (instead of by 
an experimental physicist) happens everywhere and all the time: hence 
it remains even more miraculous than before that there is a single 
outcome after each such measurement.}

\hfill Klaas Landsman, 2017 \cite[p.443]{Lan2017} 
}

\bigskip

\nopagebreak
\hfill\parbox[t]{10.8cm}{\footnotesize

{\em We need to take the existence of measurement outcomes as a priori
given, or otherwise give an account outside of decoherence of how
measurement outcomes are produced, because the property of classicality
is ultimately a statement about measurement statistics.}

\hfill Maximilian Schlosshauer, 2019 \cite[p.72]{Schl2019}
}

\bigskip

As argued in \sca{Neumaier} \cite[Section 10,5]{Neu.CQP}, decoherence
tells roughly the same story as the thermal interpretation, but only in
statistical terms, whereas the thermal interpretation refines this to a
different, more detailed story for each single case. This is possible
since in the thermal interpretation, measurement outcomes are defined 
as q-expectations of macroscopic detector quantities $X'$ strongly 
correlated -- in the way phenomenologically discussed in Section 
\ref{s.measP} -- with the microscopic quantities $X$ to be measured.
According to the thermal interpretation, a measurement $a_k$ of $X$ is
treated as an approximation of the q-expectation $\ol X$ of $X$. 
$\ol X$ is (in principle, in general only inaccurately)
\bfi{observable} if it varies sufficiently slowly with $t$ and has a 
sufficiently small uncertainty $\sigma_X$. But it may require 
considerable experimental ingenuity to do so with an uncertainty close 
to $\sigma_X$. The uncertain value $\ol X$ is considered informative 
only when its q-uncertainty $\sigma_X$ is much less than $|\ol X|$.

Since q-expectations are always single-valued, this immediately resolves
the unique outcome problem of quantum measurement theory. The thermal
interpretation makes direct sense of individual events even at the
theoretical level. It naturally gives an ontology for individual
quantum systems -- not only for thermal systems but also for microscopic
systems and even the whole universe.
In particular, because of the single-valuedness of the true values in 
the thermal interpretation, probabilities are not intrinsic to quantum 
physics but are emergent imperfections. In contrast,  Born's statistical
interpretation needs probabilities in the very foundations, due to the 
multi-valuedness of the true values.

Both interpretations are in agreement with the experimental record.
The same number $a_k$ obtained by a measurement may be interpreted in  
two ways, depending on the convention used: 
(i) It measures the q-expectation to some accuracy. 
(ii) It measures some random eigenvalue to a possibly higher (in the 
idealization even infinite) accuracy.
In both cases, the measurement involves an additional uncertainty
related to the degree of reproducibility of the measurement, given by
the standard deviation of the results of repeated measurements.
Tradition and the thermal interpretation agree in that this uncertainty
is -- by \gzit{e.Esigma} -- at least $\sigma_X$.
If the eigenvalues $X_k$ of $X$ (assumed to have discrete spectrum) are 
exactly known beforehand, one can calibrate the pointer scale to make
$a_k=X_k$ for all detector elements $k$. As long as one ignores 
the idealization error, the thermal interpretation and Born's 
interpretation become experimentally indistinguishable. However, this 
is no longer so in the more realistic case where eigenvalues are only 
approximately known -- the common situation in spectroscopy -- and must 
therefore be inferred experimentally. In this case (cf. Footnote 
${}^{\ref{f.irrat}}$), Born's statistical 
interpretation paints an inadequate, idealized picture only. The thermal
interpretation, however, still gives a correct account of the actual 
experimental situation. 
Measurements are regarded as fluctuating discrete readings of detector 
properties, defined as q-expectations of macroscopic pointer variables 
whose statistical mean agrees with the q-expectation of the system 
quantity measured, as discussed in Section \ref{s.measP}.

Thus the thermal interpretation is in full agreement with the standard 
recipes for drawing inferences from inaccurate measurement results.
The situation is precisely the same as in classical metrology, where 
observable quantities always have a true value determined by the 
theoretical description, and all randomness in measurements is assumed
to be due to measurement noise.

The thermal interpretation regards each measurement result $a_k$ as an 
approximation of the true value $\ol X$, with typical error 
$|a_k-\ol X|$ of at least $\sigma_X$, by \gzit{e.Esigma}. In the limit 
of arbitrarily many repetitions,
the statistical mean value of the approximations approaches the 
q-expectation $\ol X$, and their standard deviation approaches the
q-uncertainty $\sigma_X$. The observed discreteness is explained as an 
effect due to the recording device. The latter introduces a systematic
discretization error, of the same nature as the rounding errors in the 
illustrative example given in Subsection \ref{ss.rep}.

For example, binary responses of the macroscopic detector elements may 
be explained as in \sca{Neumaier} \cite[Chapter 11]{Neu.CQP} by a 
bistability (cf. \sca{Bonifacio \& Lugiato} \cite{BonL}, 
\sca{Gevorgyan} et al. \cite{GevSCK}) of their coarse-grained 
microscopic dynamics analogous to the bistability that give rise to 
binary responses in classical coin tossing. 
The bimodal distribution of the measurement results may be due 
to environment-induced randomness and environment-induced dissipation, 
as for a classical, environment-induced diffusion process in a 
double-well potential. For a discussion of the dynamical aspects of 
the quantum chaos responsible for this see, e.g., 
\sca{Ingraham \& Acosta} \cite{IngA},
\sca{Zhang \& Feng} \cite{ZhaF},
\sca{Belot \& Earman} \cite{BelE},
\sca{Gomez} et al. \cite{GomLL}, 
and Chapter 11 of  my book (\sca{Neumaier} \cite{Neu.CQP}).
The bimodal distribution may also be due more immediately to the 
experimental setup. For example, the arrangement in a Stern--Gerlach 
experiment together with simple theory leads to two more or less 
focussed beams, which accounts for the approximate 2-valuedness of the 
response at the screen. The thermal interpretation attributes this 
discreteness to the detection setup, not to a true, discrete value of 
the spin.

For precision measurements, e.g., those related to the quantum Hall 
effect, relevant for preparing calibration states for metrology 
standards (\sca{Lindeley} (\cite{Lin}, \sca{Kaneko} et al. 
\cite{KanNO}), one needs to be able to prepare states with tiny 
q-uncertainties, so that their measurement can be performed with very 
high accuracy. The possibility of achieving tiny uncertainties is 
linked to the spectrum. Let $X$ be a scalar or vector quantity whose
$m$ components are defined on a common dense domain.
We call the set of $\xi\in\Cz^m$ for which no linear operator $R(\xi)$ 
exists such that $R(\xi)|X-\xi|^2$ is the identity the \bfi{spectrum} 
$\spec X$\index{$\spec X$, spectrum} of $X$. The spectrum
is always a closed set, but it may be empty, as, e.g., for the vector
formed by position and momentum of a particle. The following result 
implies that the preparation of states such that $X$ has arbitrarily 
small uncertainty is possible precisely when $\ol X$ belongs to the 
spectrum of $X$. 

\begin{thm}
$\xi\in\Cz^m$ belongs to the spectrum of a scalar or vector quantity $X$
with $m$ components iff states exist that have arbitrarily small 
positive $\<|X-\xi|^2\>$.
\end{thm}

\bepf
By definition, the operator $B:=|X-\xi|^2$ is Hermitian and positive 
semidefinite, hence essentially self-adjoint with nonnegative spectrum 
$\Sigma$. Note that $\xi$ belongs to the spectrum of $X$ iff 
$0\in\Sigma$. We consider the spectral projector $P(s)$ to the invariant
subspace corresponding to the spectrum in $[0,s]$. If $0\not\in\Sigma$
then $P(s)=0$ for some $s>0$, and $\<|X-\xi|^2\>=\<B\>\ge s$ cannot be 
arbitrarily small. On the other hand, if $0\in\Sigma$ then, for all 
$s>0$, the projector $P(s)$ is nonzero. Thus we can find vectors 
$\phi(s)$ such that $\psi(s):=P(s)\phi(s)$ is nonzero. By scaling 
$\phi(s)$ appropriately we can ensure that $\psi(s)$ has norm one. 
Then in the corresponding pure state, $\<|X-\xi|^2\>=\<B\>\le s$. 
Since this works for any $s>0$, the claim follows.
\epf

Thus though in the thermal interpretation eigenvalues no longer play
the role they traditionally have in measurement, they continue to be 
essential for high precision measurements. Elsewhere in quantum physics,
they play a role primarily in the case of the Hamiltonian, where the 
eigenvalues specify the possible energy levels of a quantum system.
The latter is relevant not only on the theoretical level, where a 
spectral representation allows the explicit solution of the quantum 
Liouville equation and the time-dependent Schr\"odinger equation. 
It is also important in many contexts with experimentally relevant 
consequences: We have seen the Rydberg--Ritz formula \gzit{e.RR} from 
spectroscopy, whci can access a large number of energy levels. In many 
other cases, only few energy levels are experimentally accessible, in 
which case the quantum system can be modeled as a few 
level system, drastically simplifying the task of state tomography.
In much of quantum chemistry, only the electronic ground state is 
considered relevant. Its energy, parameterized by the nuclear 
coordinates, determines the potential energy surface whose properties 
are sufficient to describe the shape of all molecules and their 
chemical reactions. 
The energy spectrum is also prominent in thermal equilibrium physics 
since it determines the partition function from which all other 
thermodynamic properties can be derived. For a treatment of thermal 
equilibrium quantum physics in the spirit of the thermal interpretation
-- i.e., based on q-expectations without making reference to statistical
assumptions -- see the online textbook by \sca{Neumaier \& Westra} 
\cite{NeuW}.

\section{Conclusion}

In Chapter \ref{s.measP}, assuming an intuitive informal notion of 
response, a suggestive definition was given of what constitutes a 
detector. From this, the standard POVM description of general quantum 
measurements was derived,
including Born's rule in its generalized POVM form. The traditional form
of Born's rule for projective measurement followed as a special case.
Rather than postulating Born's rule for projective measurements, as done
in standard textbooks, it was derived here together with all its
ramifications and its domain of validity, from simple, easily motivated
definitions. In particular, the spectral notions appear not as
postulated input as in traditional expositions, but as consequences of
the derivation. The derivation shows that measurements satisfying Born's
rule (i.e., projective measurements) are always idealizations. Born's 
rule in textbook form does not apply when one approximately measures 
quantities with a continuous spectrum, or simultaneously measures 
quantities corresponding to noncommuting operators -- one needs a POVM 
that is not projective. Other limitations of Born's rule were pointed 
out in my recent book (\sca{Neumaier} \cite[Section 14.3]{Neu.CQP}).

Though very elementary, our notion of a detector and the associated
formal notion of a discrete POVM  is very flexible and accounts for all
basic aspects of practical measurement processes.
Apart from the simplicity and the straightforward motivation and
derivation, another advantage of the POVM setting above is that it is
absolutely clear what measurement amounts to and how accurately it may
represent given q-expectations in operational terms.

For qubits, the density operator and the dynamical equations (quantum 
Liouville, Lindblad, and Schr\"odinger equation) were derived
from classical optics. Together with the present elementary approach
to measurement, this provides a fully intelligible introduction to all
basic features of quantum mechanics.
In contrast to the usual treatments, where these basic features are
addressed by just postulating the required items, this gives actual
understanding, not only the appearance of it.
To motivate and understand Born's rule for POVMs is much easier (one
just needs simple linear algebra) than to motivate and understand
Born's rule in its original form, where unfamiliar stuff about wave
functions, probability amplitudes and spectral representations must be
swallowed by the beginner -- not to speak of the difficult notion of 
self-adjointness (which is usually simply suppressed in introductory 
treatments). Thus introductory courses on quantum mechanics would 
benefit from presenting the real thing in place of only a time-honored 
but too special, highly idealized case. An outline for a possible 
design of such a course was given in Section \ref{ss.introQM}.

In Section \ref{s.unc}, a thorough discussion was given of various 
aspects of uncertainty in quantum measurements. It culminated in a 
justification of the thermal interpretation of quantum physics, treated 
in detail in my recent book {\em Coherent quantum physics} 
(\sca{Neumaier} \cite{Neu.CQP}). This interpretation replaces the 
traditional fundamentally stochastic eigenvalue link to measurement by 
the assumption that the true properties of a quantum system are its 
q-expectations and what can be computed from these. This gives the 
stochastic aspects of measurements the same status as in classical 
mechanics, thus making the foundations of quantum physics much more 
intuitive. 

\bigskip
{\bf Acknowledgment.}
This paper benefitted from discussions with Rahel Kn\"opfel.

\bigskip

\end{document}